\documentclass[epj]{svjour}
\usepackage{graphicx}
\usepackage{amssymb}
\usepackage{amsmath}

\newcommand{\bk}{{\bf k}}
\newcommand{\bq}{{\bf q}}

\newcommand{\bK}{{\bf K}}

\newcommand{\bx}{{\bf x}}

\newcommand{\br}{{\bf r}}
\newcommand{\bR}{{\bf R}}

\newcommand{\bz}{{\bf 0}}

\renewcommand{\Im}{{\mathop{\rm{Im}}\nolimits\,}}

\newcommand{\kB}{k_{\mathrm{B}}}

\newcommand{\modified}[1]{{\relax #1}}

\title{Pairing symmetry of superconducting graphene}

\author{F. M. D. Pellegrino\inst{1,2,3} \and G. G. N.
Angilella\inst{1,2,3,4}\footnote{Corresponding author} \and R.
Pucci\inst{1,3}}
\institute{
\inst{1} Dipartimento di Fisica e Astronomia, Universit\`a di Catania,
Via S. Sofia, 64, I-95123 Catania, Italy\\
\inst{2} Scuola Superiore di Catania, Via S. Nullo, 5/i, I-95123 Catania,
Italy\\
\inst{3} CNISM, UdR Catania, I-95123 Catania, Italy\\
\inst{4} INFN, Sez. Catania, I-95123 Catania, Italy}

\abstract{
The possibility of intrinsic superconductivity in alkali-coated graphene
monolayers has been recently suggested theoretically. Here, we derive the
possible pairing symmetries of a carbon honeycomb lattice and discuss their
phase diagram. We also evaluate the superconducting local density of states
(LDOS) around an isolated impurity. This is directly related to scanning
tunneling microscopy experiments, and may evidence the occurrence of
unconventional superconductivity in graphene.
} 

\begin{document}

\maketitle

\section{Introduction}

Graphene is a one-atom-thick layer of graphite \cite{Novoselov:05a}. Due to its
nearly ideal two-dimensional character and the linear, relativistic-like
dispersion of its elementary excitations \cite{CastroNeto:08}, graphene is an
intriguing material, where correlation and reduced dimensionality may conspire
in favor or against various electronic instabilities. Indeed, it has been
proposed that an electron liquid in a honeycomb lattice can be characterized by 
several ordered states, depending on doping and on the electron-electron
interaction \cite{Honerkamp:08}. Among these competing orders, superconductivity
(SC) could be stabilized by either topological disorder \cite{Gonzalez:01a} or
the proximity to an electronic topological transition (ETT) \cite{Gonzalez:08}.
In this case, the symmetry of the underlying lattice may allow for an
unconventional structure of the SC order parameter, and possibly sizeable
critical temperatures $T_c$, as is the case for the high-$T_c$ cuprates
\cite{BlackSchaffer:07,Jiang:08,Baskaran:02}. 

The great interest in future graphene-based technology largely owes to the
relatively high and easily tunable conductivity that can be realized in clean
graphene samples. Electronic correlations and the possible occurrence of ordered
phases are therefore often neglected. Recently, a superconducting current has
been observed to propagate through a superconductor-normal-superconductor (SNS)
Josephson junction, where the N region consisted of a graphene layer
\cite{Heersche:07}. This provided evidence of SC phase coherence in graphene
single layers. Other carbon-based compounds are also known to sustain
superconductivity, sometimes with fairly large critical temperatures $T_c$.
These include (a) highly oriented pyrolytic graphite (HOPG)
\cite{Kopelevich:00}, (b) the graphite intercalated compounds (GIC)
\cite{Dresselhaus:02,Csanyi:05,Weller:05}, which may be described as graphene
sheets alternated by alkali layers, mainly acting as charge reservoirs, (c)
quasi-one-dimensional carbon nanotubes, and (d) quasi-zero-dimensional
alkali-doped fulleres, or fullerides \cite{Gunnarsson:97}. In particular, recent
\emph{ab initio} calculations support the idea that in-plane phonons may be
responsible of SC in GIC
\cite{Calandra:06,Kim:06a,Boeri:07,Sanna:07,Liu:07a,Grueneis:09}. The quest for
(quasi)intrinsic superconductivity in quasi-two-dimensional graphene is
therefore well motivated, and several mechanisms have been proposed. These range
from conventional, phonon-mediated superconductivity as in the GIC
\cite{Calandra:06,Kim:06a,Boeri:07,Sanna:07,Liu:07a,Grueneis:09,Valla:09}, to
unconventional, electronic mechanisms, including particularly the resonating
valence bond (RVB) mechanism \cite{Baskaran:02,BlackSchaffer:07}. Moreover, it
has been proposed \cite{Uchoa:07a} that SC may be realized in alkali-coated
graphene single layers, where either electronic states belonging to the metallic
bands get paired by means of graphene phonon modes, or graphene electrons couple
via metallic plasmons \cite{Uchoa:07a}.

Here, after classifying the symmetries of a SC order parameter compatible with
the honeycomb lattice of graphene, and discuss their stability and possible
mixing as a function of doping and coupling strengths, we study the local
density of states (LDOS) of SC graphene around an isolated impurity
\cite{Wehling:08b}. In particular, we suggest that scanning tunneling microscopy
(STM) measurements can distinguish among the various available SC symmetries in
graphene, thereby evidencing the possible unconventional nature of the SC order
parameter, as is the case for the superconducting state of high-$T_c$ cuprates
\cite{Hudson:99,Pan:99}, and possibly of their unconventional normal state
\cite{Chakravarty:01,Morr:02,Andrenacci:04c}.

\section{Superconducting phase diagram}
\label{sec:SC}

We start by considering a model Hamiltonian $H=H_0 + H_1$, where $H_0 =
\sum_{\bk\lambda} \xi_{\bk\lambda} c^\dag_{\bk\lambda} c_{\bk\lambda}$ describes
the normal electron liquid, with $c^\dag_{\bk\lambda}$ ($c_{\bk\lambda}$) a
creation (annihilation) operator for a quasiparticle with wavevector $\bk$
within the first Brillouin zone (1BZ) and band index $\lambda=\pm$,
$\xi_{\bk\lambda} = E_{\bk\lambda} -\mu$ the tight-binding dispersion relation
for band $\lambda$, measured with respect to the chemical potential $\mu$.
Retaining hopping and overlap terms between nearest neighbor sites
\cite{Saito:98}, one has $E_{\bk\lambda} = \lambda t |\gamma_\bk|/(1-\lambda
s|\gamma_\bk|)$, where $t=2.8$~eV and $s=0.07$ are the nearest neighbor hopping
and overlap parameters, respectively \cite{Reich:02}, and $\gamma_\bk =
\sum_{\ell=1}^3 e^{i\bk\cdot\delta_\ell}$ is the usual (complex) structure
factor in momentum space. Here, $\delta_1 = a(1,\sqrt{3})/2$, $\delta_2 =
a(1,-\sqrt{3})/2$, $\delta_3 = a(-1,0)$ are the vectors connecting nearest
neighbor sites in real space, where $a=0.142$~nm is the C--C distance
\cite{CastroNeto:08}.

\begin{figure}[tb]
\begin{center}
\begin{minipage}[c]{0.48\columnwidth}
\begin{center}
\includegraphics[height=\textwidth,angle=-90]{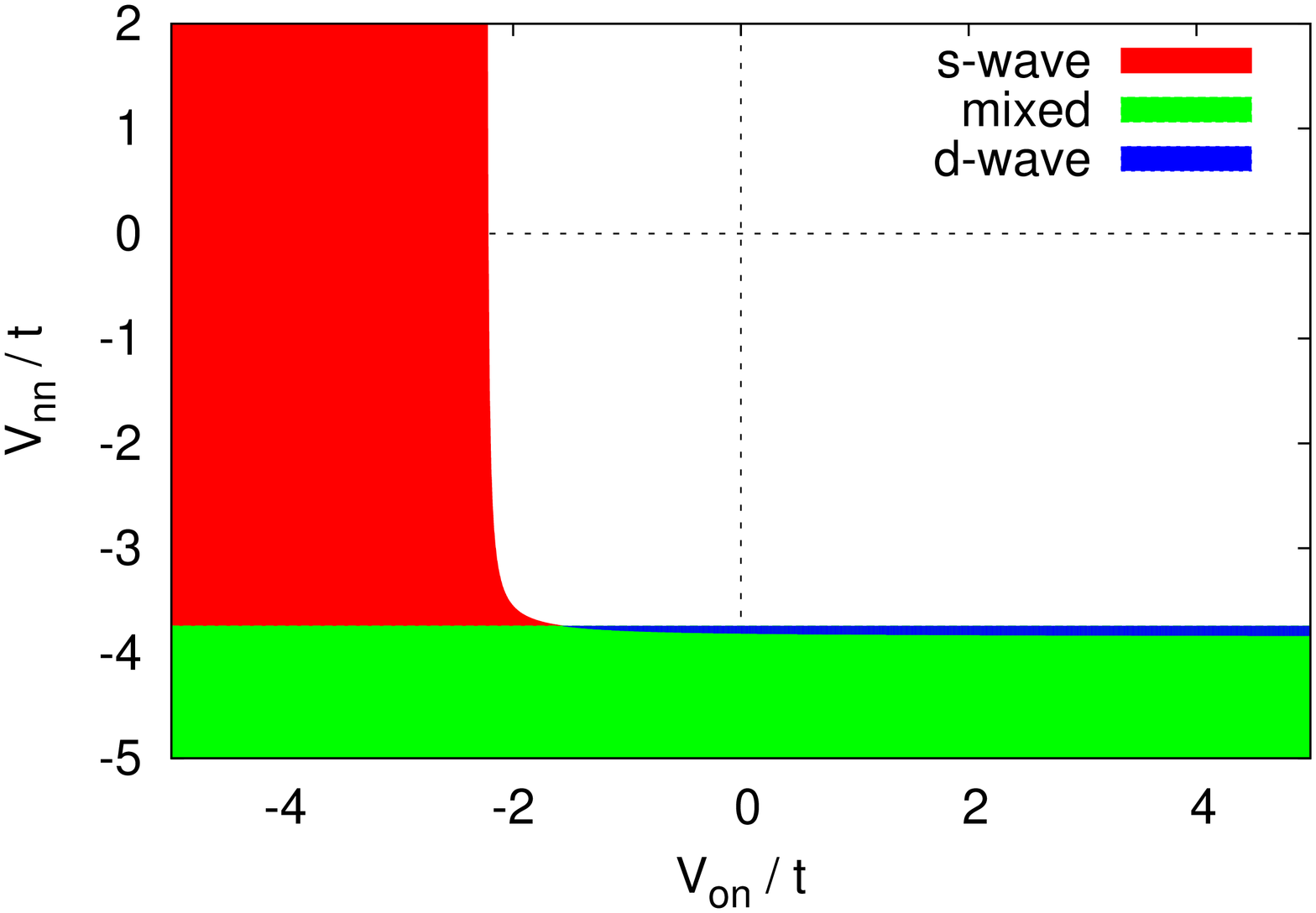}
\end{center}
\end{minipage}
\begin{minipage}[c]{0.48\columnwidth}
\begin{center}
\includegraphics[height=\textwidth,angle=-90]{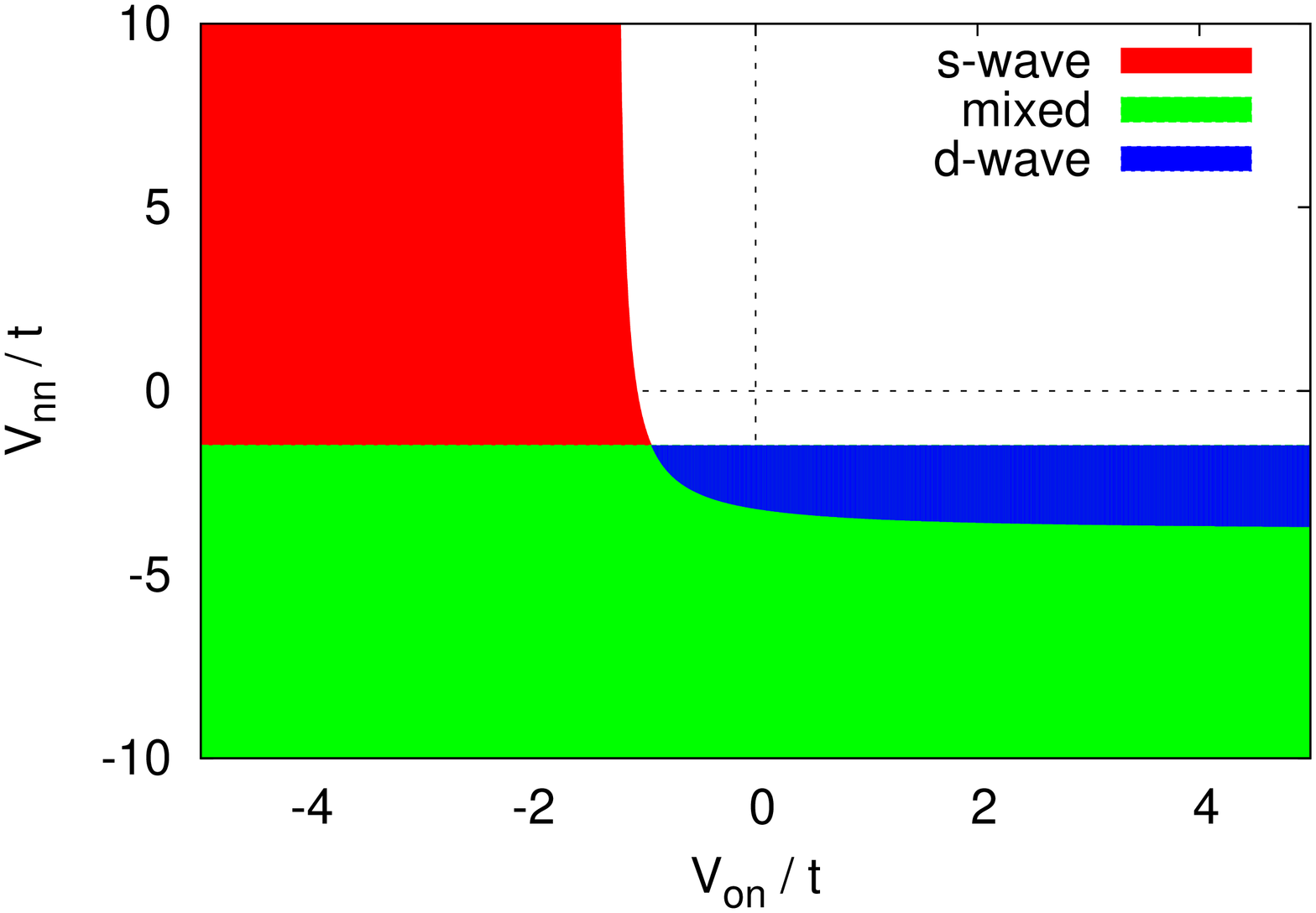}
\end{center}
\end{minipage}
\begin{minipage}[c]{0.48\columnwidth}
\begin{center}
\includegraphics[height=\textwidth,angle=-90]{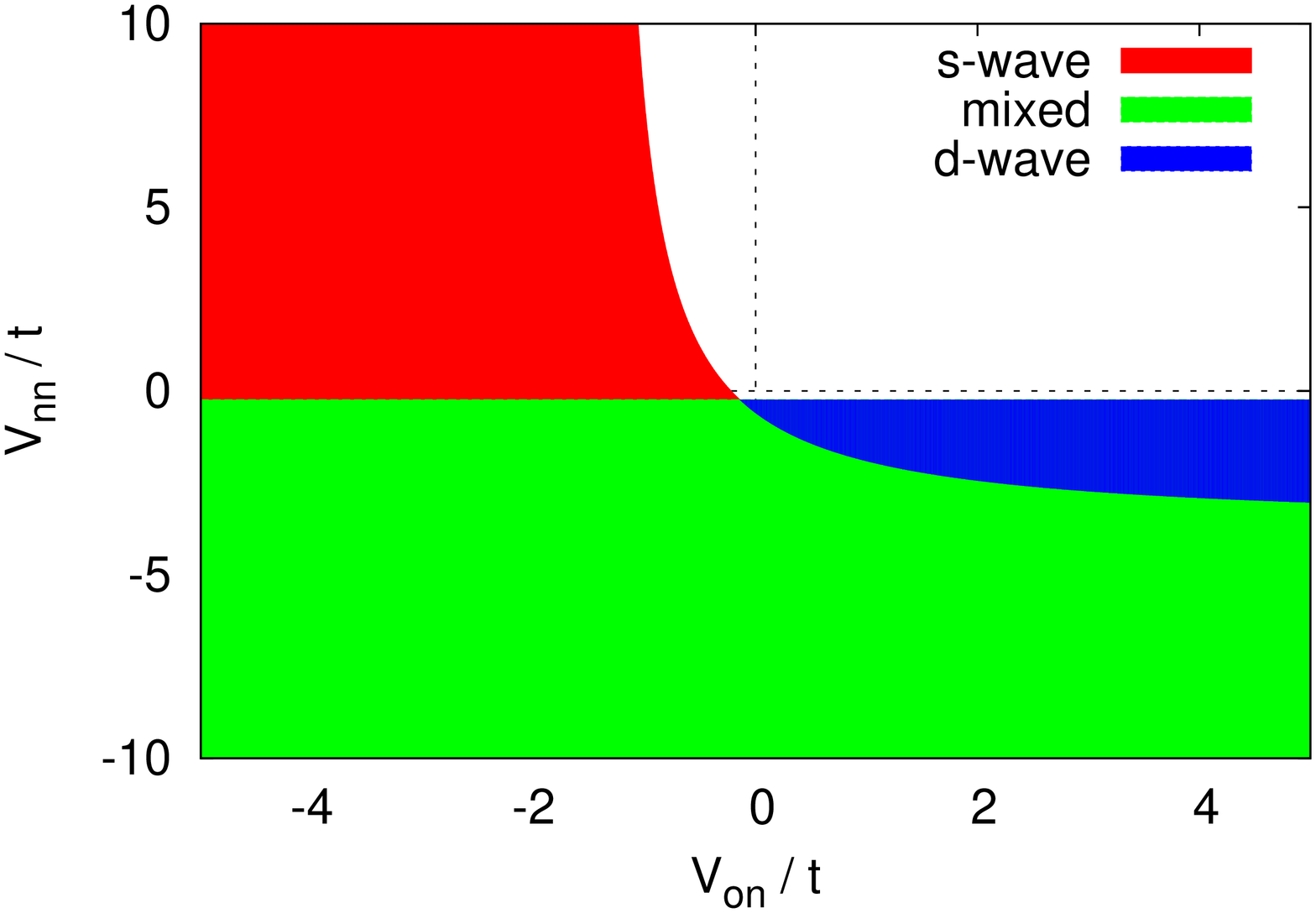}
\end{center}
\end{minipage}
\begin{minipage}[c]{0.48\columnwidth}
\begin{center}
\includegraphics[height=\textwidth,angle=-90]{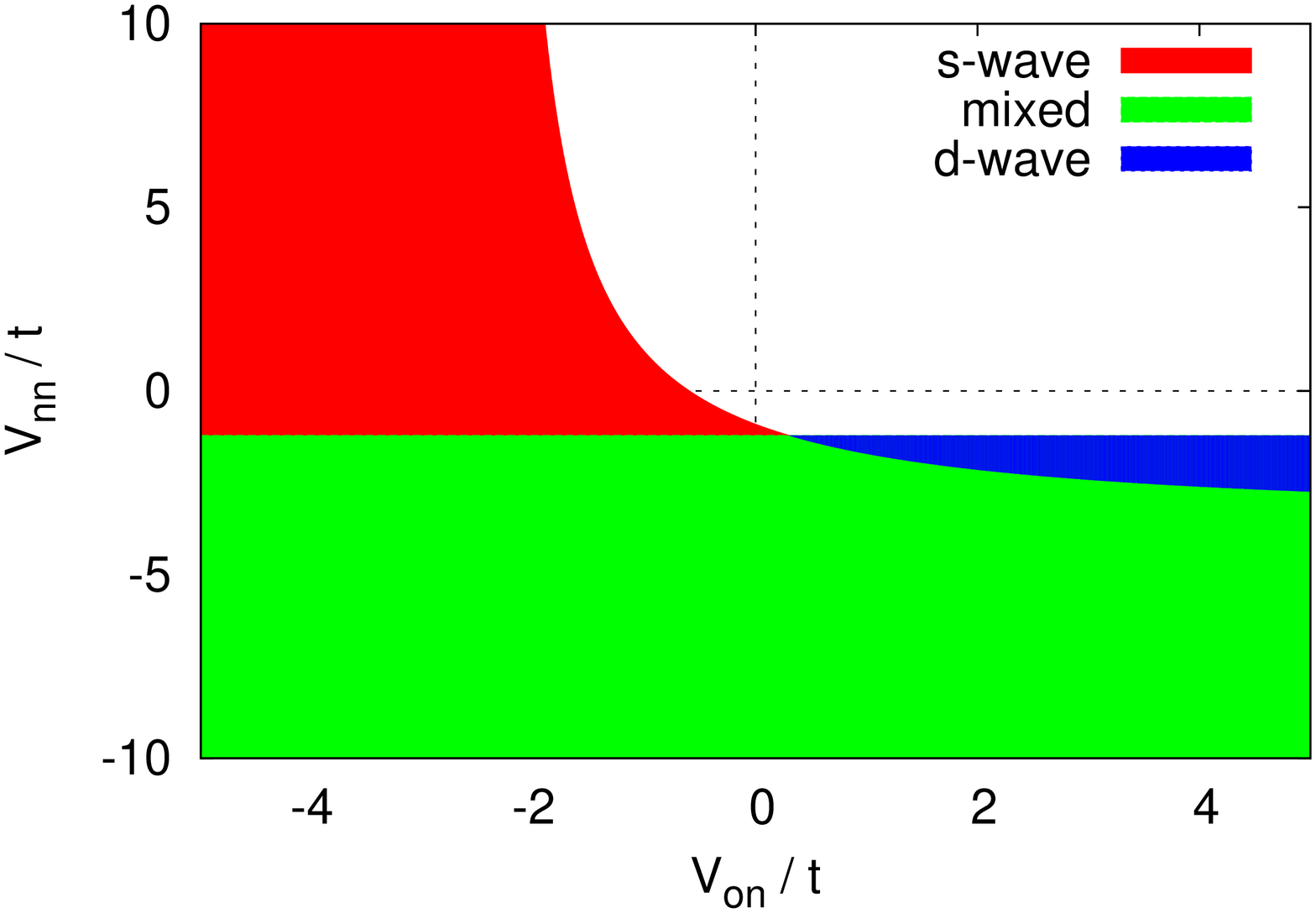}
\end{center}
\end{minipage}
\end{center}
\caption{(Color online) Low-temperature phase diagrams for the symmetry of the
SC order parameter as a function of the coupling strengths. Left to right, top
to bottom panels show the cases $\mu=0$ (undoped), $0.5t$, $1.07t$ (Van~Hove
singularity), and $1.5t$.}
\label{fig:diagrams}
\end{figure}

As for the pairing Hamiltonian $H_1$, we restrict to onsite ($V_{on}$) and
nearest-neighbor ($V_{nn}$) interaction only. In terms of separate sets of creation
and annihilation operators $a^\dag_{\bk\sigma}$, $a_{\bk\sigma}$
($b^\dag_{\bk\sigma}$, $b_{\bk\sigma}$) for the $A$ ($B$) sublattices, along the
singlet channel and within the mean-field approximation, one finds
\begin{eqnarray}
H_1 &=& \sum_\bk \Delta_0 (a^\dag_{\bk\uparrow} a^\dag_{-\bk\downarrow} +
b^\dag_{\bk\uparrow} b^\dag_{-\bk\downarrow}) \nonumber\\
&&+ \Delta_1 (\bk)
(a^\dag_{\bk\uparrow} b^\dag_{-\bk\downarrow} - a^\dag_{\bk\uparrow}
b^\dag_{-\bk\downarrow} ) + \mathrm{H.c.},
\end{eqnarray}
where the two components of the order parameter are defined as
\begin{subequations}
\begin{eqnarray}
\Delta_0 &=& \frac{V_{on}}{N} \sum_{\bk^\prime} \langle a_{-\bk^\prime\downarrow}
a_{\bk^\prime\uparrow} \rangle = \frac{V_{on}}{N} \sum_{\bk^\prime} \langle b_{-\bk^\prime\downarrow}
b_{\bk^\prime\uparrow} \rangle ,\\
\Delta_1 (\bk ) &=& \frac{V_{nn}}{2N} \sum_{\bk^\prime} \gamma_{\bk-\bk^\prime}
\langle b_{-\bk^\prime \downarrow} a_{\bk^\prime \uparrow} - b_{-\bk^\prime
\uparrow} a_{\bk^\prime \downarrow} \rangle .
\end{eqnarray}
\label{eq:gaps}
\end{subequations}
While the onsite order parameter $\Delta_0$ is manifestly $\bk$-independent, the
presence of an anisotropic intersite component $\Delta_1 (\bk)$ opens the
possibility of unconventional superconductivity. In particular, it is possible
to factorize $\Delta_1 (\bk )$ in terms of the basis functions of the
irreducible representations (irreps) of the point group $D_{6h}$ as
\begin{equation}
\Delta_1 (\bk ) = \sum_{\ell=0}^2 \Delta_{1\ell} [\phi_\ell (\bk) + i \psi_\ell
(\bk)],
\end{equation}
where $\phi_0 (\bk) + i\psi_0 (\bk) = \gamma_\bk / \sqrt{3}$ and  $\phi_\ell
(\bk) + i \psi_\ell (\bk) = e^{i\bk\cdot\delta_\ell} -\gamma_\bk/3$
($\ell=1,2,3$) \cite{Pellegrino:10}.
Their subsets $\{\phi_0\}$, $\{\psi_0\}$, $\{\phi_1, \phi_2\}$, $\{ \psi_1 ,
\psi_2 \}$ form a basis for the irreps $A_{1g}$, $B_{1u}$, $E_{2g}$, and
$E_{1u}$, respectively. One finds that the basis functions of both
one-dimensional irreps vanish at the Dirac points $\bK = \frac{2\pi}{3a}
(1,\frac{\sqrt{3}}{3} )$, $\bK^\prime = \frac{2\pi}{3a}
(1,-\frac{\sqrt{3}}{3})$. 
Following standard nomenclature \cite{Jiang:08}, a nonvanishing component
$\Delta_0$, $\Delta_{01}$, and $\Delta_{0\ell}$ ($\ell=1,2$) will be termed
$s$-wave, extended $s$-wave, and $d$-wave, respectively. We have analyzed the
gap equations (\ref{eq:gaps}) as a function of temperature and chemical
potential, as well as of the coupling constants. As is well known, when several
competing symmetries are available for the order parameter, the gap equations
factorize at $T_c$, and symmetry mixing is only possible below $T_c$
\cite{Angilella:99}. Since the system is actually two-dimensional, the
mean-field $T_c$ only provides an upper bound for the
Berezinskii-Kosterlitz-Thouless transition \cite{Fossheim:04}.
Fig.~\ref{fig:diagrams} shows the low-temperature mean-field phase diagrams for
graphene, at various dopings, as a function of the coupling strengths.  
\modified{%
Solid lines separating filled regions are defined as the locus in the plane of
reduced coupling constants, $(V_{on}/t,V_{nn}/t)$, such that the system exhibits
SC with the relatively low critical temperature $T_c = 4$~K. The limiting case
of pristine graphene ($\mu=0$, Fig.~\ref{fig:diagrams}, upper left panel)
requires relatively high coupling strengths in order to develop SC even with a
critical temperature as low as $T_c =4$~K. At $\mu=0$, as $T\to0$, one
recover the phase diagram of Ref.~\cite{Uchoa:07a}, where however only the $s$-
and $p+ip$-waves symmetries have been addressed. When $\mu\neq0$, one finds that $T_c
\to0$ when $V_{on},V_{nn}\to0^-$, in agreement with results of
Refs.~\cite{Uchoa:07a,Uchoa:05,Zhao:06}.}
One then needs finite doping in order to
have SC at moderately low coupling strengths (Fig.~\ref{fig:diagrams}, upper
right panel, $\mu=0.5t$). Upon doping, $d$-wave symmetry prevails at $T_c$ close
to the Van~Hove singularities in both the valence and conduction bands, while
$s$- or extended $s$-wave symmetry wins out close to the band edges. This is a
generic effect of the proximity to an ETT, and has been described as a feature
of high-$T_c$ superconductivity in the cuprates \cite{Angilella:01}. At exactly
the ETT within the conduction band (Fig.~\ref{fig:diagrams}, lower left panel,
$\mu=1.07t$), $d$-wave symmetry is maximally favored. It has been suggested
\cite{Gonzalez:08} that fine-tuning the gate voltage in suspended graphene may
achieve dopings of practical use. In particular, this may be even easier in
uniaxially strained graphene, where the energy slope of the DOS increases with
increasing strain \cite{Pellegrino:09b}. Upon further increasing the chemical
potential (Fig.~\ref{fig:diagrams} lower right panel, $\mu=1.5t$), the pure
$d$-wave SC region shrinks, and eventually gets suppressed
altogether. Assuming $V_{on}=0$ ($\Delta_0 = 0$), and neglecting band asymmetry
($s=0$), one recovers the results of \cite{BlackSchaffer:07,Jiang:08}. In
particular, $V_{nn} = -2.4t$ corresponds to Baskaran's proposal of resonating
bond ordering in graphite \cite{Baskaran:02}.

\section{Single impurity effects}
\label{sec:impurity}

We next include the effect of a nonmagnetic localized impurity, whose
potential is given by
\begin{equation}
V=a^2 U_0 \sum_\sigma \Psi^\dag_\sigma (\bx) \Psi_\sigma (\bx) ,
\end{equation}
where $\Psi_\sigma (\bx)$ is a field operator at position $\bx$ with spin
projection $\sigma$, and the energy $U_0$ measures the impurity potential
strength. In terms of the Nambu spinor $\varphi_\bk^\dag = \left(
a^\dag_{\bk\uparrow} , b^\dag_{\bk\uparrow} , a_{-\bk\downarrow} ,
b_{-\bk\downarrow} \right)$, the imaginary time Green's function is defined as
$\mathcal{G} (\bk,\bk^\prime ,\tau) = - \langle \varphi_\bk (\tau)
\varphi_{\bk^\prime}^\dag (0)\rangle$, and obeys the Gor'kov matrix equation of
motion
\begin{equation}
\sum_\bq [(i\omega_n - \mathcal{K}_\bk ) \delta_{\bk\bq} - \mathcal{V}(\bk,\bq)]
\mathcal{G}(\bq,\bk^\prime,i\omega_n ) = \delta_{\bk\bk^\prime} ,
\label{eq:Gorkov}
\end{equation}
where $\hbar\omega_n = (2n+1)\pi\kB T$ is a fermionic Matsubara frequency. Here,
$\mathcal{K}_\bk$ is a $4\times4$ matrix associated with the mean-field SC
Hamiltonian,
\begin{equation}
\mathcal{K}_\bk = \begin{pmatrix} K_\bk^0 & \Delta_\bk \\ \Delta_\bk^\dag & -
K_\bk^0 \end{pmatrix} ,
\end{equation}
whose $2\times2$ blocks are given by
\begin{subequations}
\begin{eqnarray}
K_\bk^0 &=& \begin{pmatrix} \frac{1}{2} (E_{\bk-} + E_{\bk+})-\mu &
(E_{\bk-}-E_{\bk+})/(2\gamma_\bk^\ast) \\
(E_{\bk-}-E_{\bk+})/(2\gamma_\bk)  & \frac{1}{2} (E_{\bk-} + E_{\bk+})-\mu
\end{pmatrix} ,\\
\Delta_\bk &=& \begin{pmatrix} \Delta_0 & \Delta_1 (\bk) \\ \Delta_1 (-\bk) &
\Delta_0 \end{pmatrix} .
\end{eqnarray}
\end{subequations}
Analogously, the impurity potential enters as a $4\times4$ block-diagonal matrix
\begin{equation}
\mathcal{V}(\bk,\bk^\prime) = \begin{pmatrix} V(\bk,\bk^\prime) & 0 \\ 0 &
-V(\bk,\bk^\prime ) \end{pmatrix} ,
\end{equation}
whose $2\times2$ blocks have generic elements
\begin{equation}
V_{\lambda\lambda^\prime} (\bk,\bk^\prime) = a^2 U_0 \psi_{\bk\lambda}^\ast
(\bx) \psi_{\bk^\prime\lambda^\prime} (\bx) ,
\end{equation}
and $\psi_{\bk\lambda}(\bx)$ is the Bloch wavefunction for sublattice
$\lambda=A,B$ employed in the tight-binding diagonalization of the pure sector
of the Hamiltonian \cite{Pellegrino:09}. Eq.~(\ref{eq:Gorkov}) is exactly
soluble in the SC case, in the absence of impurities, while in the normal case
and in the presence of localized impurities at highly-symmetric lattice
positions it has been studied in \cite{Pellegrino:09}. In the general case, its
solution can be expressed in terms of an appropriate $T$-matrix as
\begin{eqnarray}
\mathcal{G}(\bk,\bk^\prime,i\omega_n) &=& \delta_{\bk\bk^\prime} \mathcal{G}_0
(\bk,i\omega_n) \nonumber\\
&&\hspace{-1truecm}+ \mathcal{G}_0 (\bk,i\omega_n) T(\bx;\bk,\bk^\prime,i\omega_n )
\mathcal{G}_0 (\bk^\prime ,i\omega_n)
\end{eqnarray}
where $\mathcal{G}_0 (\bk,i\omega_n)$ is the solution of the SC pure case.
Fourier transforming back in real space via $\psi_{\bk\lambda}(\br)$ and
performing an analytic continuation to real frequencies, the local density of
states at position $\br$ can be expressed as $\rho(\br,\omega)=-\frac{1}{\pi}\Im
G(\br,\br,\omega)$. In what follows, we will use Gaussian pseudoatomic
wavefunctions $\phi(\br)=Z_g \exp(-Z_g^2 r^2/24a^2)/(2a\sqrt{3\pi})$
($Z_g=11.2$) \cite{Pellegrino:09} to expand the Bloch wavefunctions within the
tight-binding approximation as $\psi_{\bk\lambda}(\br)=N^{-1/2}\sum_j
\phi(\br-\bR_j^\lambda)e^{i\bk\cdot\bR_j^\lambda}$, where $\bR_j^\lambda$ are
vectors of the $\lambda=A,B$ sublattices. Placing the impurity on an $A$ site,
at $\br=\bz$ say, and retaining only the zeroth order
approximation, one has $\psi_{\bk A}(\bz)\approx N^{-1/2} \phi(\bz)$ and
$\psi_{\bk B} (\bz) \approx 0$. In this limit, the impurity potential matrix
simplifies to $\mathcal{V}(\bk,\bk^\prime) \equiv \mathcal{V}_0 = \phi^2 (\bz)
a^2 U_0 N^{-1} \sigma_3 \otimes (\sigma_0 + \sigma_3)/2$, where $\sigma_0$,
$\sigma_3$ are Pauli matrices. Correspondingly, the analytically continued
$T$-matrix assumes the $\bk$-independent form $T(\br=\bz;\omega) = \mathcal{V}_0
\mathcal{M}^{-1} (\bz,\omega)$, where zeroes of the determinant of $\mathcal{M}
(\bz,\omega) = 1 - \sum_\bq G_0 (\bq,\omega) \mathcal{V}_0$ as a function of
$\omega$ are connected with impurity-induced bound states or well-defined
resonances.

\begin{figure}[tb]
\centering
\includegraphics[height=0.8\columnwidth,angle=-90]{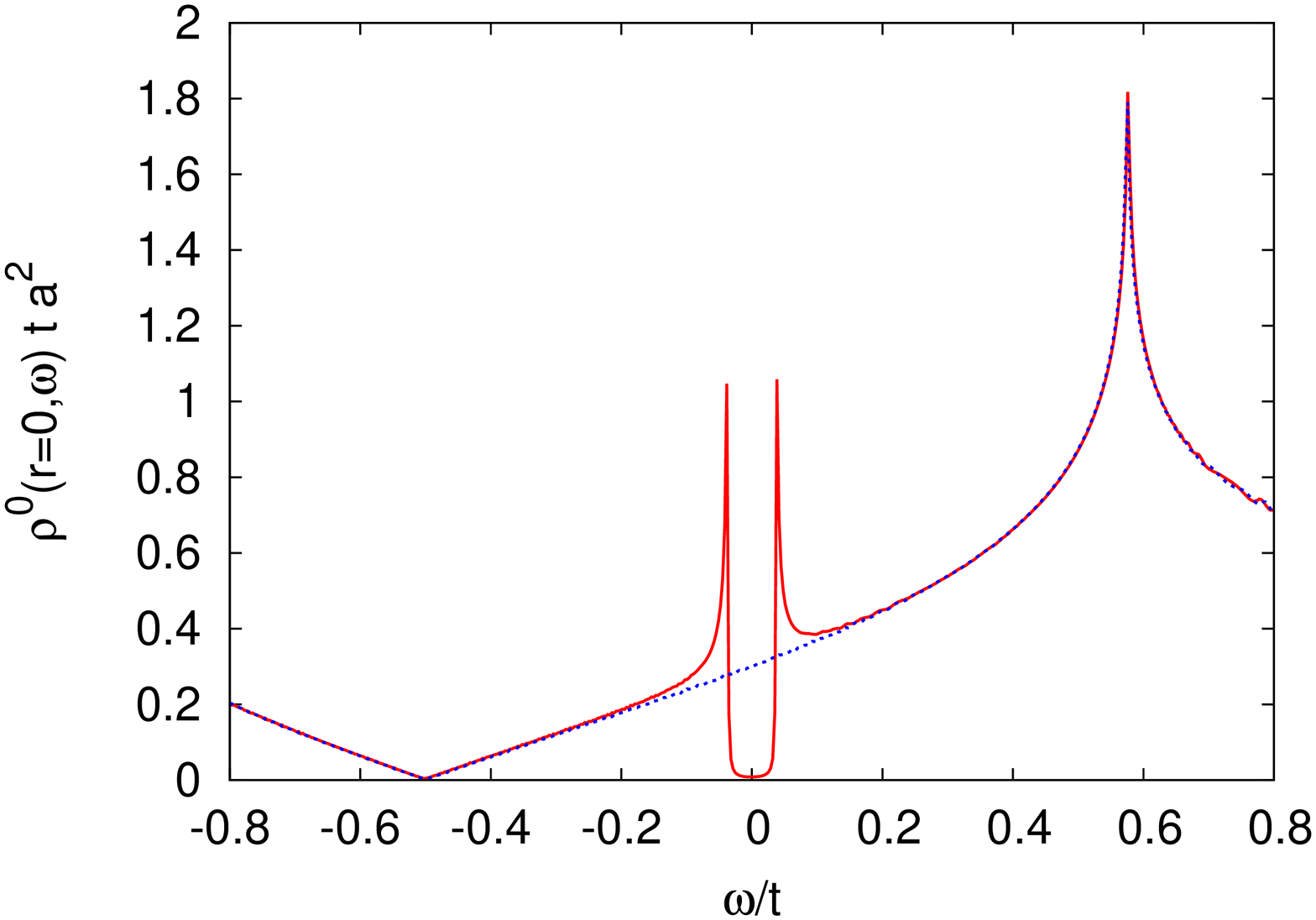}
\includegraphics[height=0.8\columnwidth,angle=-90]{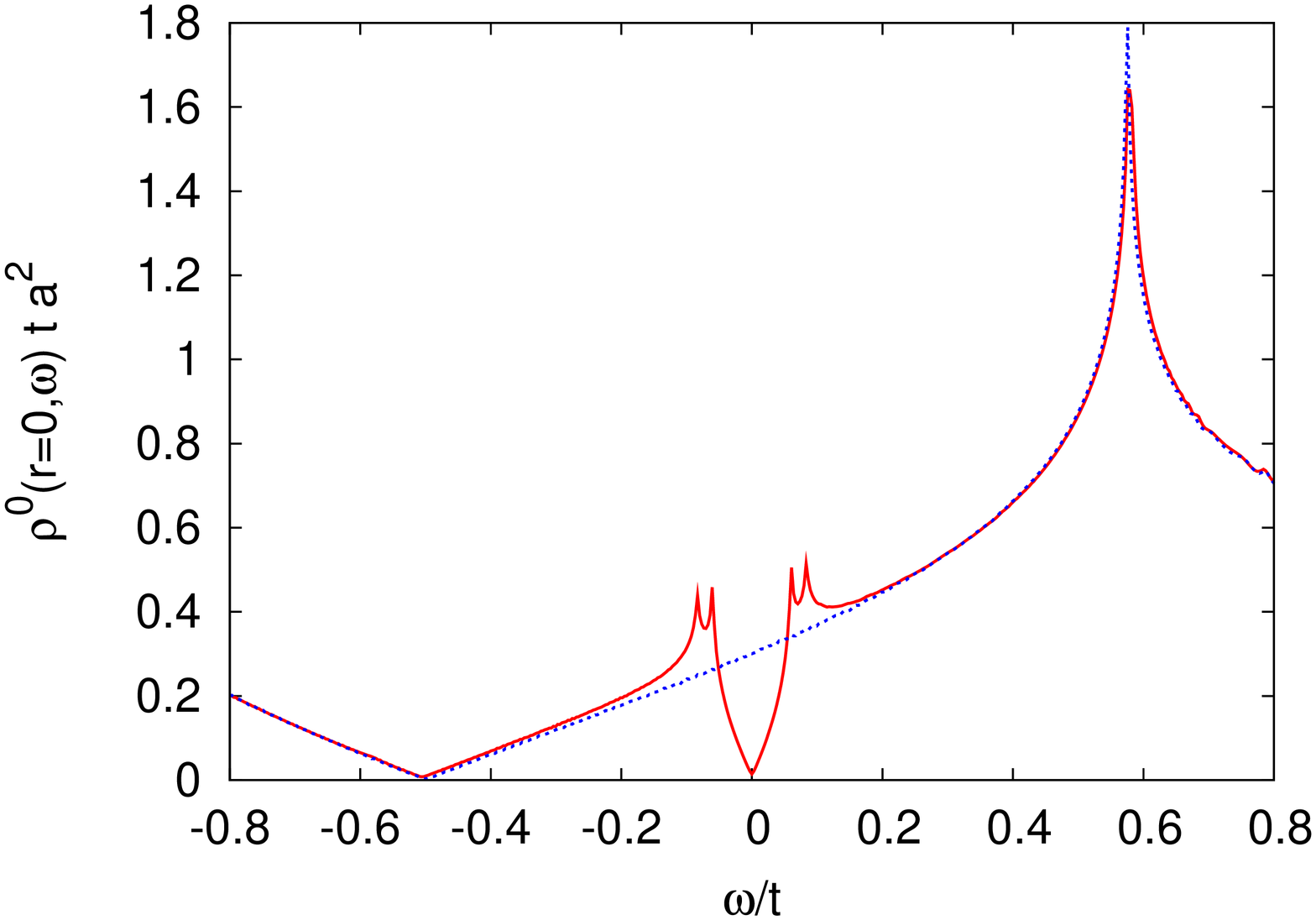}
\caption{(Color online) LDOS for pure SC (solid line) and normal-state graphene
(dotted line). Top panel refers to $s$-wave symmetry, while bottom panel refers
to $d$-wave symmetry, with in-phase components of the order parameter. In both
cases, $\mu=0.5t$ and $U_0 = 0$.}
\label{fig:ldos1}
\end{figure}

Fig.~\ref{fig:ldos1} shows the LDOS on an atomic site ($\br=\bz$) for pure SC
graphene ($U_0 =0$). Here, and in what follows, we shall consider slightly doped
graphene ($\mu=0.5t$), in order to stabilize the SC phase without requiring
exceedingly high coupling strengths. This corresponds to having the Fermi
surface in the shape of two disconnected rings, centered around either
inequivalent Dirac point. In the $s$-wave case (Fig.~\ref{fig:ldos1}, top
panel), the LDOS assumes a typical BCS-like shape, with well-defined coherence
peaks located at $\omega\approx\pm\Delta_0 /2$, and no spectral weight is
available within the gap. Away from the gapped region, the LDOS is not affected
by superconductivity, and in particular the Van~Hove singularities remain
intact. Qualitatively similar results have been obtained in the case of an
extended $s$-wave order parameter. This is because, despite the nontrivial
$\bk$-dependence, the order parameter vanishes precisely at the Dirac points,
\emph{viz.} where the bands vanish if $\mu=0$. On
the other hand, in the $d$-wave case (Fig.~\ref{fig:ldos1}, bottom panel), the
onsite LDOS vanishes linearly at $\omega=0$, and the coherence peaks are
considerably reduced. This is typical of unconventional superconductors, and is
in fact a hallmark of $d$-wave superconductivity in the cuprates \cite{Morr:02}.
In Fig.~\ref{fig:ldos1}, we have assumed in-phase $d$-wave components for the
order parameter, \emph{i.e.} $\Delta_{11}=\Delta_{12}$. Other phase relations
are possible, such as $\Delta_{12}=\Delta_{11} e^{-i\pi/3}$. In the latter case,
$\Delta_1(\bk)$ vanishes only at the Dirac points. Therefore, in the case of
doped graphene, the SC order parameter has no node along the Fermi surface, and
one finds a fully gapped LDOS, qualitatively similar to the $s$-wave case. This
is in agreement with the results of \cite{Jiang:08}. However, while an out-of-phase
order parameter is numerically favored in the case of pure $d$-wave symmetry, we
find that a nonzero $s$-wave component stabilizes in-phase $d$-wave
superconductivity.

\begin{figure}[tb]
\centering
\includegraphics[height=0.8\columnwidth,angle=-90]{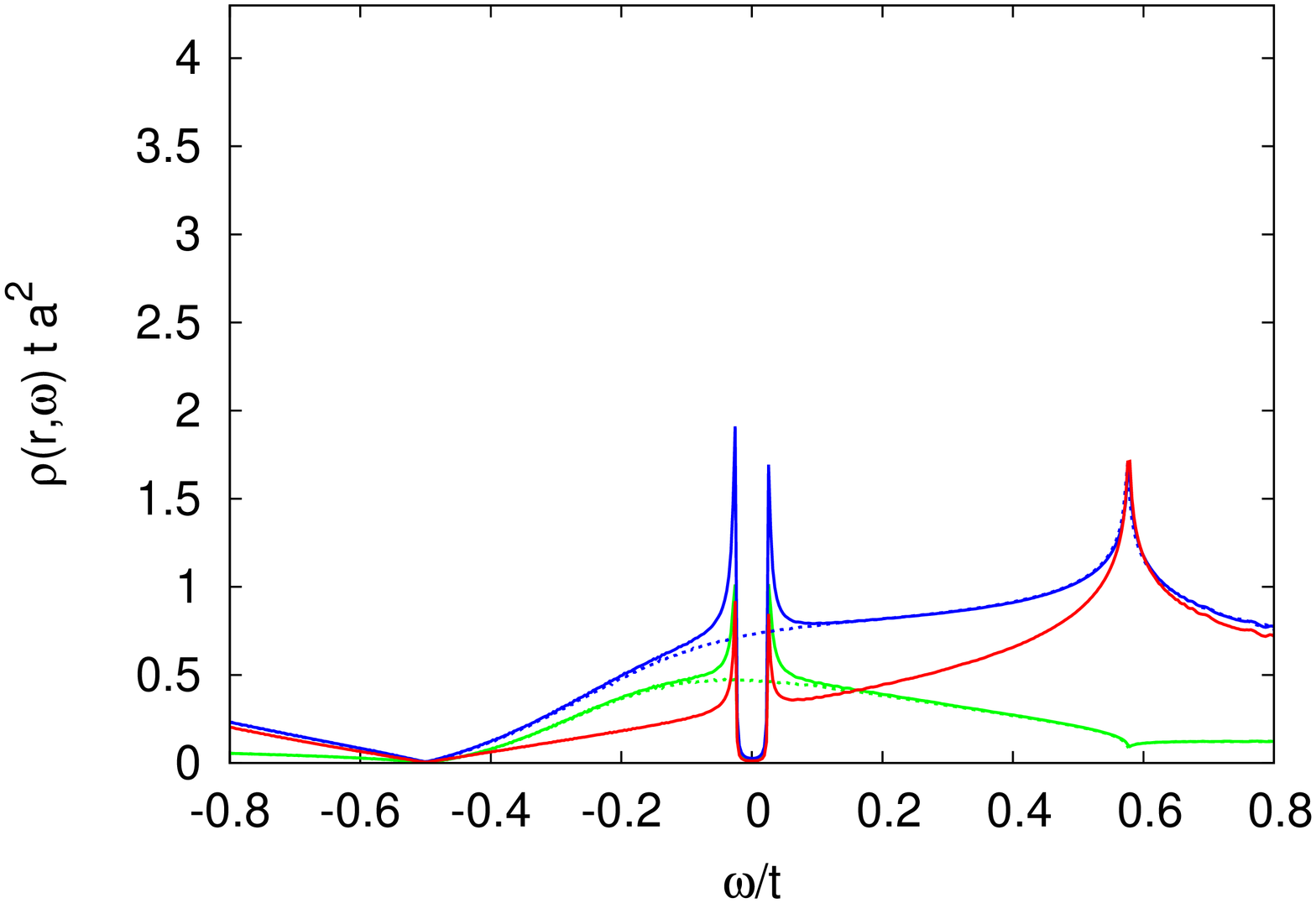}
\includegraphics[height=0.8\columnwidth,angle=-90]{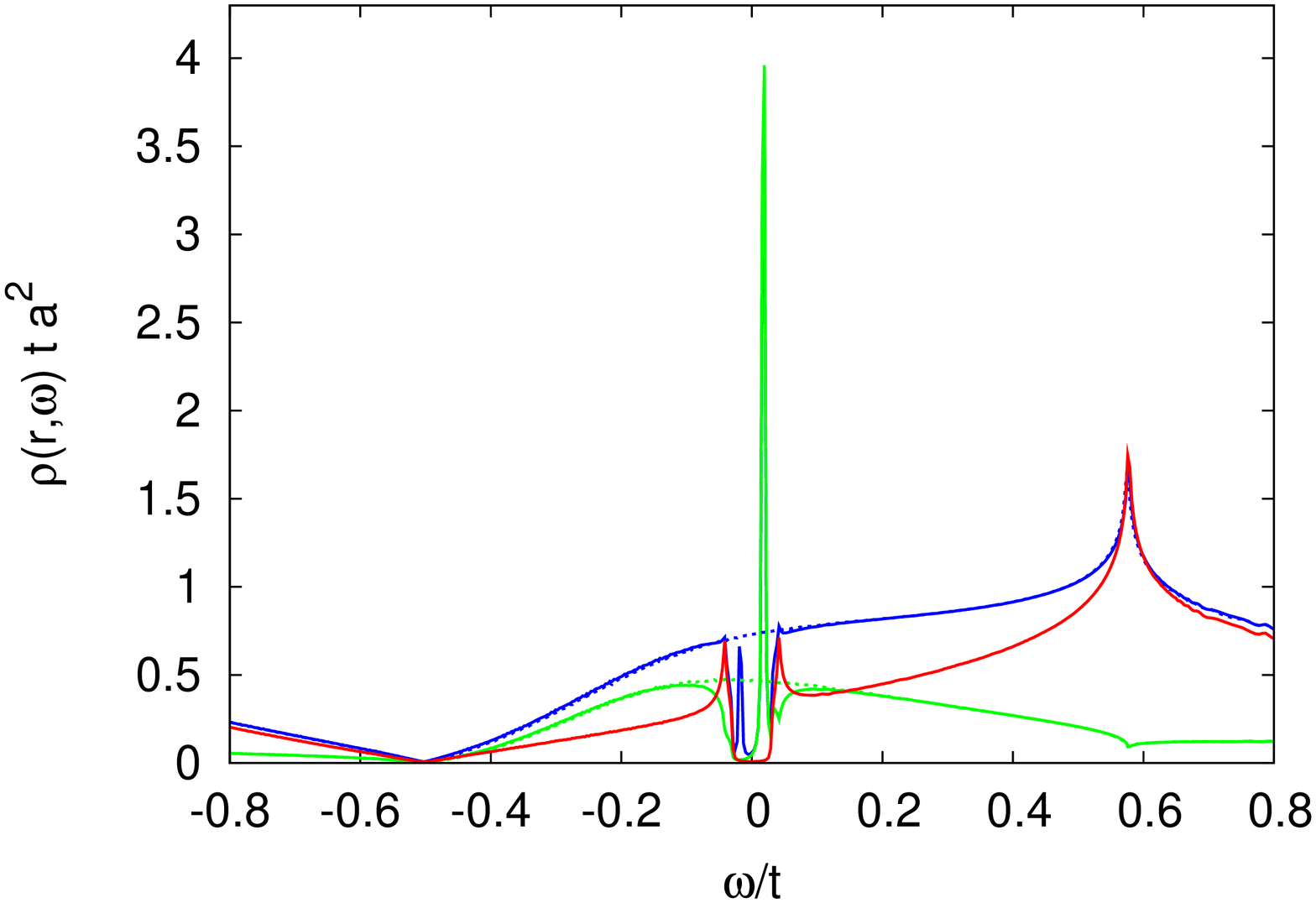}
\caption{(Color online) LDOS in the presence of a localized impurity for SC
(solid lines) and normal-state graphene (dotted lines). Green lines refer to
LDOS on top of the impurity ($\bx=\bz$), while blue lines refer to LDOS on a
nearest-neighbor site ($\bx=\delta_3$). Red lines refer to the on-site LDOS
($\bx=\bz$) in the SC phase, in the absence of impurities. Top panel refers to
extended $s$-wave symmetry, while bottom panel refers to $d$-wave symmetry. In
both cases, $\mu=0.5t$ and $U_0 = -0.75t$.}
\label{fig:ldos2}
\end{figure}

We consider now the effect of a single, localized, nonmagnetic impurity, located
on a lattice site. In the normal state, we recover the results of
\cite{Pellegrino:09} for a site-like impurity. Fig.~\ref{fig:ldos2} shows our
results for the LDOS in the SC state. In the $s$-wave case
(Fig.~\ref{fig:ldos2}, top panel), the onsite LDOS is characterized by the
opening of a gap and the formation of well-defined coherence peaks, while the
Van~Hove singularity gets suppressed, as in the normal state. On the other hand,
the LDOS on a nearest-neighbor site gets enhanced with respect to its normal
state counterpart, and the Van~Hove singularities undergo no significant
reduction. More importantly, by analyzing the determinant of the matrix
$\mathcal{M}(\br,\omega)$, one finds that no bound state is possible within the
gapped region in the $s$-wave case. In the $d$-wave case (Fig.~\ref{fig:ldos2},
bottom panel), while a similar analysis applies as in the $s$-wave case at large
energies, the situation is completely different within the gapped region. One
finds that sufficiently high impurity coupling strengths ($-3.8t\lesssim U_0
\lesssim -0.8t$, in the case under study) allow the formation of well-defined
bound state pairs within the gapped region. (That bound states should come in
pairs follows from the particle-hole mixing, characteristic of the SC state
\cite{Morr:02}.) This is indeed apparent from Fig.~\ref{fig:ldos2} (bottom
panel), with the appearance of sharp peaks in the onsite LDOS. Such a result is
similar to the $d$-wave SC state of the cuprates, and has indeed been verified
experimentally in STM experiments in Bi2212 \cite{Hudson:99,Pan:99}.

\section{Conclusions}
\label{sec:conclusions}

In summary, we derived the possible symmetries of a SC order parameter
compatible with the honeycomb lattice structure of graphene. We discussed the SC
phase diagram at low-temperature as a function of doping and interaction
strengths. In particular, we showed that unconventional pairing may stabilize
close to an ETT, with possible admixtures of a subdominant $s$-wave
contributions. 

We then evaluated the LDOS around an isolated impurity in the SC phase, and
suggested that STM experiments may detect the occurrence of unconventional
pairing in intrinsically or proximity-induced superconducting graphene. This
will be helpful in determining the pairing mechanism of superconducting
graphene. We have limited our study to isolated nonmagnetic impurities. Such a
condition may be realized in the limit of low impurity concentration, where it
is safe to neglect impurity effects on the overall transport properties of
graphene.

\begin{small}
\bibliographystyle{epj}
\bibliography{a,b,c,d,e,f,g,h,i,j,k,l,m,n,o,p,q,r,s,t,u,v,w,x,y,z,zzproceedings,Angilella}
\end{small}

\end{document}